\documentclass[aps, pre, onecolumn, superscriptaddress, floatfix, arxiv, 11pt]{revtex4-2}
\usepackage{amsmath, amssymb, graphicx, bm}
\usepackage{hyperref}
\hypersetup{colorlinks,allcolors=blue}

\begin{document}
\title{Shape-dependent motility of polar inclusions in active baths}

\author{Pritha Dolai}
\email{pritha.dolai@fau.de}
\thanks{Equal contribution}
\affiliation{Friedrich-Alexander-Universit\"at, Erlangen-N\"urnberg, Germany 91054 }
\affiliation{International Centre for Theoretical Sciences, Tata Institute of Fundamental Research,
Bengaluru, India 560089}

\author{Aditya Singh Rajput}
\email{aditya.rajput@icts.res.in}
\thanks{Equal contribution}

\author{K. Vijay Kumar}
\email{vijaykumar@icts.res.in}
\affiliation{International Centre for Theoretical Sciences, Tata Institute of Fundamental Research,
Bengaluru, India 560089}

\date{\today}

\begin{abstract}
Collections of persistently moving active particles are an example of a nonequilibrium heat bath. One way to study the nature of nonequilibrium fluctuations in such systems is to follow the dynamics of an embedded probe particle. With this aim, we study the dynamics of an anisotropic inclusion embedded in a bath of active particles. By studying various statistical correlation functions of the dynamics, we show that the emergent motility of this inclusion depends on its shape as well as the properties of the active bath. We demonstrate that both the decorrelation time of the net force on the inclusion and the dwell time of bath particles in a geometrical trap on the inclusion have a non-monotonic dependence on its shape. We also find that the motility of the inclusion is optimal when the volume fraction of the active bath is close to the value for the onset of motility induced phase separation.
\end{abstract}

\maketitle

\section{Introduction}

The study of systems driven out of equilibrium by a throughput of energy at the level of the individual units has revealed a rich set of nonequilibrium collective states \cite{ramaswamyMechanicsStatisticsActive2010, marchettiHydrodynamicsSoftActive2013,bechingerActiveParticlesComplex2016,fodorStatisticalPhysicsActive2018}. The physical realization of such active materials range from molecular motors inside cells driven by ATP to mechanically agitated granular materials. Active particles which form the basic constituents of these materials are typically anisotropic in shape with associated orientational degrees of freedom. The hydrodynamic theory of such active liquid crystals has been a topic of great interest in the past decades \cite{marchettiHydrodynamicsSoftActive2013, julicherHydrodynamicTheoryActive2018}. 

A complementary direction of research considers isotropic particles wherein the relationship between dissipation and noise in their dynamics is not in accordance with the fluctuation-dissipation theorem \cite{tailleurStatisticalMechanicsInteracting2008,catesWhenAreActive2013}. Examples of such systems include Janus particles with asymmetric surface reactions, spherical droplets with asymmetric internal flows, and driven isotropic granular particles. Such active particles form the basis of scalar active matter \cite{wittkowskiScalarF4Field2014}. Three commonly studied models of active particles are the Run-and-Tumble Particles (RTPs), Active Brownian Particles (ABPs), and Active Ornstein-Uhlenbeck Particles (AOUPs). An emergent property seen in systems of such particles with purely repulsive interactions is a nonequilibrium ``condensed" phase arising due to the interplay between activity and density. This motility-induced-phase-separation (MIPS) is, in fact, seen in homogeneous systems of RTPs, ABPs and AOUPs \cite{tailleurStatistical2008, catesMotilityInducedPhaseSeparation2015,filyAthermal2012, filyFreezing2014, fodorHow2016}.

Interacting systems of active particles provide examples of nonequilibrium heath baths and studying the dynamics of probes embedded in such reservoirs provide insight into the statistical properties of active baths. For instance, the dynamics of tagged active particles have recently been observed to display universal behaviors across the three models mentioned above both in a single-file geometry \cite{dolaiUniversalScalingActive2020} and in harmonic chains
\cite{singhCrossoverBehavioursExhibited2021}. Another approach would be to embed tracers or inclusions in active heat baths, and follow their dynamics. Systems comprised of active particles with embedded isotropic passive inclusions can have depletion forces that can be either attractive or repulsive \cite{dolaiPhaseSeparationBinary2018,NajiDepletionPRE2020,NajiSciRep2020}. Essentially, the accumulation of active particles near boundaries changes the nature of the depletion forces between passive colloidal particles \cite{bechingerActiveParticlesComplex2016}. It is well appreciated that these fluctuation-induced depletion forces are sensitive to the shape of the inclusions in equilibrium systems \cite{israelachvili2011intermolecular}. Such depletion forces also arise in active baths \cite{aminovFluctuationInducedForcesNonequilibrium2015, baekGenericLongRangeInteractions2018, deblaisBoundariesControlCollective2018}. However, for passive inclusions embedded in active heat baths, the role of geometry of embedded objects has not received much attention \cite{granekAnomalous2022, wuTransport2018,SheaSM2022,ChengExptSM2022,SiminSM2020,malloryCurvatureinducedActivationPassive2014,Milos2020, tracerDynamicsMeanField2021,angelaniGeometricallyBiasedRandom2010}. 

In this study, we explore the emergent dynamics of a passive polar inclusion embedded in a nonequilibrium active bath. In particular, we focus on how the shape of the inclusion affects its macroscopic behavior. Generically, fore-aft asymmetry and an energy flux can lead to net currents in many-body systems. Does a polar inclusion in an active heat bath display persistent motion? How does the persistent velocity of the inclusion depend on its shape? How does the interaction between shape polarity and activity translate to emergent motility of the inclusion? Active particles are known to accumulate near static boundaries \cite{dharRTP2019}. In fact, static wedge-shape objects are known to trap active particles moving around them \cite{kaiserHowCaptureActive2012, kaiserCapturingSelfpropelledParticles2013}. How does the accumulation of the active bath particles around a polar and \emph{movable} inclusion lead to its emergent motility? How do the parameters of the active bath affect the movement of the embedded inclusion? Our results are as follows: (i) we find emergent persistent dynamics of the inclusion in active baths composed of RTPs, ABPs and AOUPs, (ii) there is an optimum shape of the inclusion that leads to an enhanced motility; the correlations of the net force on the inclusion and the typical dwell time of the active particles in a trap around the inclusion have a similar non-monotonic dependence on the inclusion geometry, (iii) the motility of the inclusion is controlled by the volume fraction of the active particles in the heat bath; in particular, the motility is optimal for a volume fraction close to the onset of MIPS in the heat bath.

The paper is organised as follows. In section \ref{sec 2: model}, we discuss the construction of the polar inclusion and the parameters that characterize its shape, and present the governing dynamical equations for both the inclusion and the active particles that make up the heat bath. Next in section \ref{sec 3: results}, we describe our results on the emergent motility of the inclusion and its dependence on geometry. We end with a summary of our study and an appropriate discussion in section \ref{sec4: discussion and conclusion}.

\section{Model}\label{sec 2: model}
\begin{figure}[t]
\includegraphics[width=0.9\linewidth]{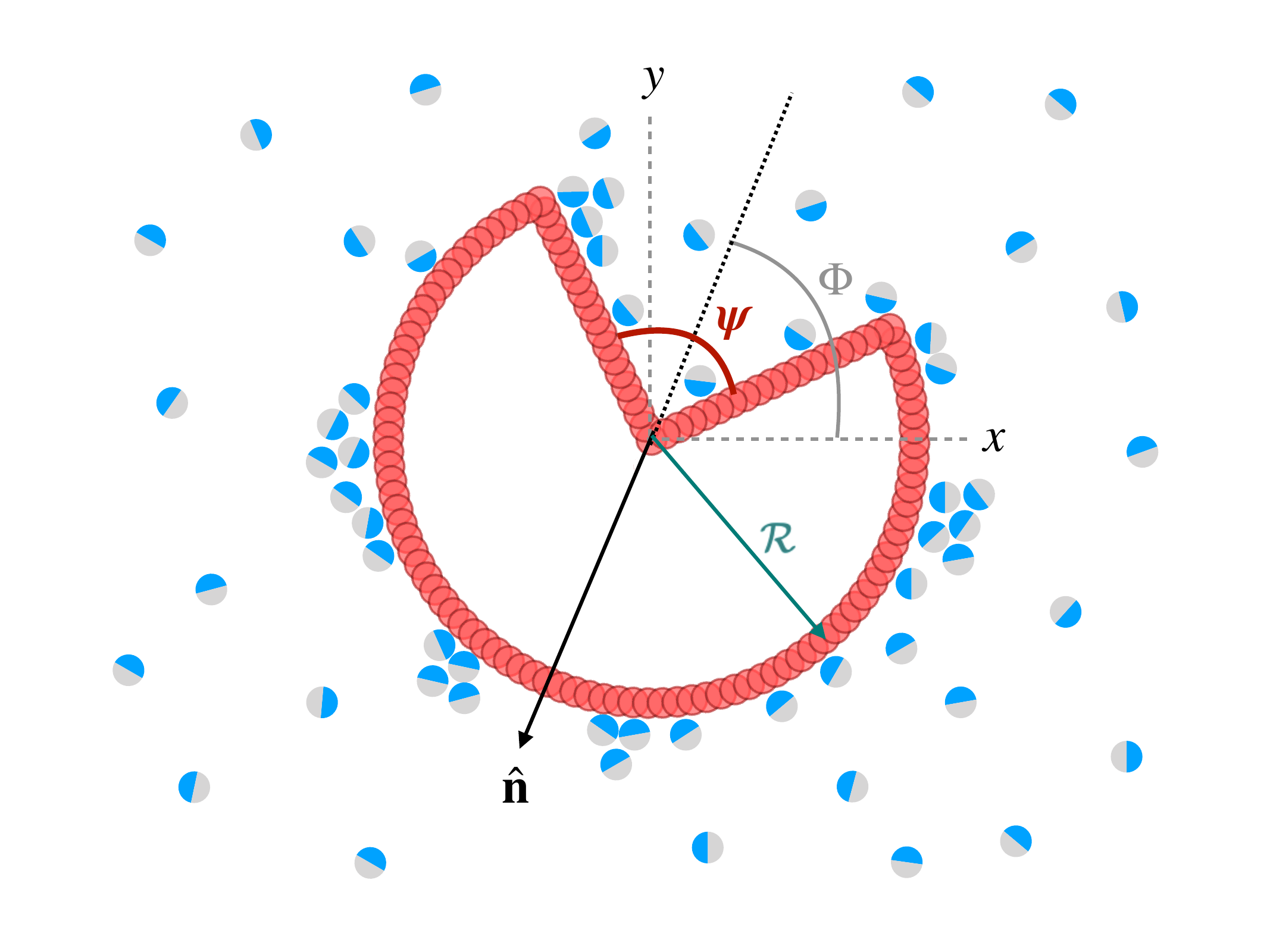}
\caption{
A polar inclusion in an active heat bath. The anisotropic inclusion is constructed from $N_I$ particles rigidly connected to each other in the geometry shown schematically. The inclusion, with opening angle $\psi$ and the radius $\mathcal{R}$ defining its overall shape, has its center-of-mass at $\mathbf{R}$ and its orientation is given by the unit normal $\hat{\mathbf{n}} = -(\cos\Phi \, \hat{\mathbf{x}} + \sin\Phi \, \hat{\mathbf{y}})$ where $\Phi$ is the angular orientation. The nonequilibrium heat bath is composed of active particles either of the RTP, ABP or AOUP kind. These active particles (whose instantaneous direction of self-propulsion is indicated by the blue/grey semicircles) exert forces on the embedded inclusion due to their persistent motion. Note that the active particles interact amongst themselves and also with the particles that make up the inclusion via the same repulsive WCA potential \eqref{eq:wca}.} 
\label{fig:schematic}
\end{figure}

We study a passive polar inclusion embedded in an active nonequilibrium ``heat bath'' with a distinct fore-aft asymmetry as shown in FIG. \ref{fig:schematic}. The impacts of the active particles on the inclusion are the driving force for its dynamics. The inclusion is constructed by rigid bonds between $N_{I}$ particles placed in an appropriate geometry. The nonequilibrium heat bath consists of $N_{B}$ scalar active particles with short-range repulsive interactions between themselves and also with the particles that constitute the inclusion. Without loss of generality, we choose the purely repulsive interaction to be the Weeks-Chandler-Anderson (WCA) potential
\begin{align}
U(r)=\epsilon \left\{ \begin{array}{lr}  \frac{1}{4}+ \left(\frac{\sigma}{r}\right)^{12}- \left(\frac{\sigma}{r}\right)^{6}, &  r<a
\\ 0, & r>a,
\end{array}
\right.
\label{eq:wca}
\end{align}
where $\epsilon$ and $\sigma$ are the characteristic energy and length scales of the potential, and $a=2^{1/6}\sigma$ is the interaction range which also determines the effective size of the particles. The overdamped dynamics of $\mathbf{r}_i$, the position $i^{\rm th}$ active particle is
\begin{align}
\frac{d{\mathbf{r}}_i}{dt} &= {\mathbf{v}}_i 
- \mu \left[ \sum_{j=1, \, j\neq i}^{N_{B}} \nabla_{\mathbf{r}_i} U({\mathbf{r}}_i-{\mathbf{r}}_j) 
+ \sum_{p=1}^{N_{I}} \nabla_{\mathbf{r}_i} U(\mathbf{x}_{p}-\mathbf{r}_i) \right]
\nonumber \\ & \qquad
+ \sqrt{2 \mu k_B T} \, {\boldsymbol{\eta}}_i(t)
\label{eq:EOM_bath_particle}
\end{align}
where $\mathbf{v}_i$ is the active self-propulsion velocity, $\mu$ is the translational mobility, and  ${\boldsymbol{\eta}}_i(t)$ is a Gaussian white noise with zero mean and unit variance. In the above equation, $\mathbf{x}_p$ is the position of the $p^{\rm th}$ particle making up the inclusion where $p=1,2,\ldots N_{I}$. The rigid inclusion is characterised by its center of mass position $\mathbf{R}$ and its orientation $\Phi$ [see FIG. \ref{fig:schematic}]. As such, $\mathbf{R}$ and $\Phi$ evolve in response to the net force and torque applied on the inclusion from the impacts of the bath particles. The overdamped evolution equations for these quantities are
\begin{align}
\frac{d\mathbf{R}}{dt} & = \mathcal{M}_t \; \mathbf{F}(t) + \sqrt{2 \mathcal{M}_t k_B T} \; \boldsymbol{\mathcal{N}}(t) 
\label{eq:incusion_position}
\\[1em]
\frac{d\Phi}{dt} &=  \mathcal{M}_r \, \mathsf{T}(t) 
+ \sqrt{2\mathcal{M}_r k_B T} \; \mathcal{Z}(t) \label{eq:incusion_angle}
\end{align}
where $\mathcal{M}_t$ ($\mathcal{M}_r$) is the translational (rotational) mobility of the inclusion, while $\boldsymbol{\mathcal{N}}(t)$ and $\mathcal{Z}(t)$ are Gaussian white noises with zero mean and unit variance, and the force $\mathbf{F}$ and the torque $\mathsf{T}$ on the center-of-mass of the inclusion due to the impacts of the bath particles are
\begin{align}
\mathbf{F}&= -\sum_{p=1}^{N_{I}} \sum_{i=1}^{N_B} \nabla_{\mathbf{x}_p} U(\mathbf{x}_{p}-\mathbf{r}_i),
\label{eq:force_on_inclusion}
\\
\mathsf{T}&=-\hat{\mathbf{z}} \cdot \sum_{p=1}^{N_{I}} \sum_{i=1}^{N_B} (\mathbf{x}_{p} - \mathbf{R}) \times \nabla_{\mathbf{x}_p} U(\mathbf{x}_{p}-\mathbf{r}_i).
\label{eq:torque_on_inclusion}
\end{align}
Notice that if the active noise $\mathbf{v}_i=0$, then the equations \eqref{eq:EOM_bath_particle}, \eqref{eq:incusion_position} and \eqref{eq:incusion_angle} represent the dynamics of a polar inclusion in an equilibrium heat bath. In this case, both the bath particles and the inclusion will have the Boltzmann distribution as their steady-state probability. However, when the bath-particles are active, the polar shape of the inclusion can lead to non-trivial emergent dynamics. Before we discuss this dynamics, we first describe the kind of active particles that we consider in this study.

The active nature of the nonequilibrium heat bath arises from the persistent self-propelled motion of its constituent particles. We consider three kinds of active baths: those comprised of (1) run-and-tumble particles (RTPs), (2) active Brownian particles (ABPs), and (3) active Ornstein-Uhlenbeck particles (AOUPs). These three scalar active particle models differ in the nature of the active stochastic forces that propel them. Specifically, the active velocity of the $i^{\rm th}$-RTP is given by $\mathbf{v}^{\rm RTP}_i = v_{\rm R} \, (\cos \theta_i, \sin \theta_i)$, where the direction of self-propulsion $0 \leq \theta_i < 2\pi$ tumbles at a Poisson rate $\gamma$ and $v_{\rm R}$ is the speed during the active runs. In other words, the orientation $\theta_i$ of each RTP changes abruptly and stochastically after a mean run-time $1/\gamma$. On the other hand, the 
internal orientation 
$\varphi_i$ of the $i^{\rm th}$-ABP performs rotational diffusion and also governs the instantaneous active velocity
\begin{align}
\frac{d\varphi_i}{dt} = \sqrt{2D_r} \; \xi_{i}(t),
\qquad
\mathbf{v}^{\rm ABP}_i = v_{\rm A} \, (\cos \varphi_i, \sin \varphi_i)
\end{align}
where $v_{\rm A}$ is the self-propulsion speed, $D_r$ is the rotational diffusion constant and $\xi_i(t)$ is a Gaussian white noise with zero mean and unit variance. Thus the orientation of the ABPs evolves continuously in time. Finally, the active velocity of AOUPs is governed by the following equation
\begin{align}
\tau \frac{d{\mathbf{v}}^{\rm AOUP}_i}{dt}=-{\mathbf{v}}^{\rm AOUP}_i + \sqrt{2\Delta} \; \zeta_i(t)
\end{align}
where $\tau$ is the persistence time of self-propelled motion, $\Delta$ is the noise strength and $\zeta_i(t)$ is a Gaussian white noise. A characteristic active speed for AOUPs can be defined as $v_{\rm O}=\sqrt{\Delta/\tau}$. Note that, the mean-squared-displacement (MSD) for free active particles (of the three kinds discussed above) have the same universal form \cite{dolaiUniversalScalingActive2020}. All three active models can be characterized by a generalized active speed $u$ and a persistence rate $\omega$. One can easily show that $u=v_R=v_A/\sqrt{2}=v_O$ and $\omega=\gamma=D_r/2=1/(2\tau)$. 

\section{Results \label{sec 3: results}}
We study the system of active particles and anisotropic passive inclusion with an opening angle $\psi$ and a radius $\mathcal{R}$ in a two-dimensional periodic box of side $L$. The schematic of our model system is shown in FIG. \ref{fig:schematic}. The inclusion has different shapes depending upon its opening angle $\psi$. The active bath is characterized by a P\`eclet number $\textrm{Pe} = u/ (a \omega)$ and an area fraction $\phi=(N_B \pi a^2 + A_{\rm inc})/L^2$ where we approximate the area enclosed by the inclusion $A_{\rm inc} \approx (\pi-\psi/2) \mathcal{R}^2$. We choose $\sigma$, $\sigma^2 / \mu \epsilon$ and $\epsilon/\sigma$ respectively as the units of length, time and force. The equations \eqref{eq:EOM_bath_particle}, \eqref{eq:incusion_position} and \eqref{eq:incusion_angle} together with the  dynamical equations for the bath particles are numerically integrated using the Euler-Maruyama scheme with a non-dimensional time step $\Delta t=10^{-3}$. We fixed the thermal energy at $k_BT/\epsilon = 0.01$, the radius of the inclusion at $\mathcal{R}/\sigma = 5$ (unless specified), and the linear size of the simulation domain at $L/\mathcal{R} = 10$. All simulations are run for long times ($\sim 150\,\omega^{-1}$) and various statistical averages and correlations, denoted by $\langle \ldots \rangle$, are computed after the system reaches a steady-state, and the errorbars are calculated over several realizations.

\subsection{Emergent active dynamics of the polar inclusions}

\begin{figure}[ht]
\includegraphics[width=\linewidth]{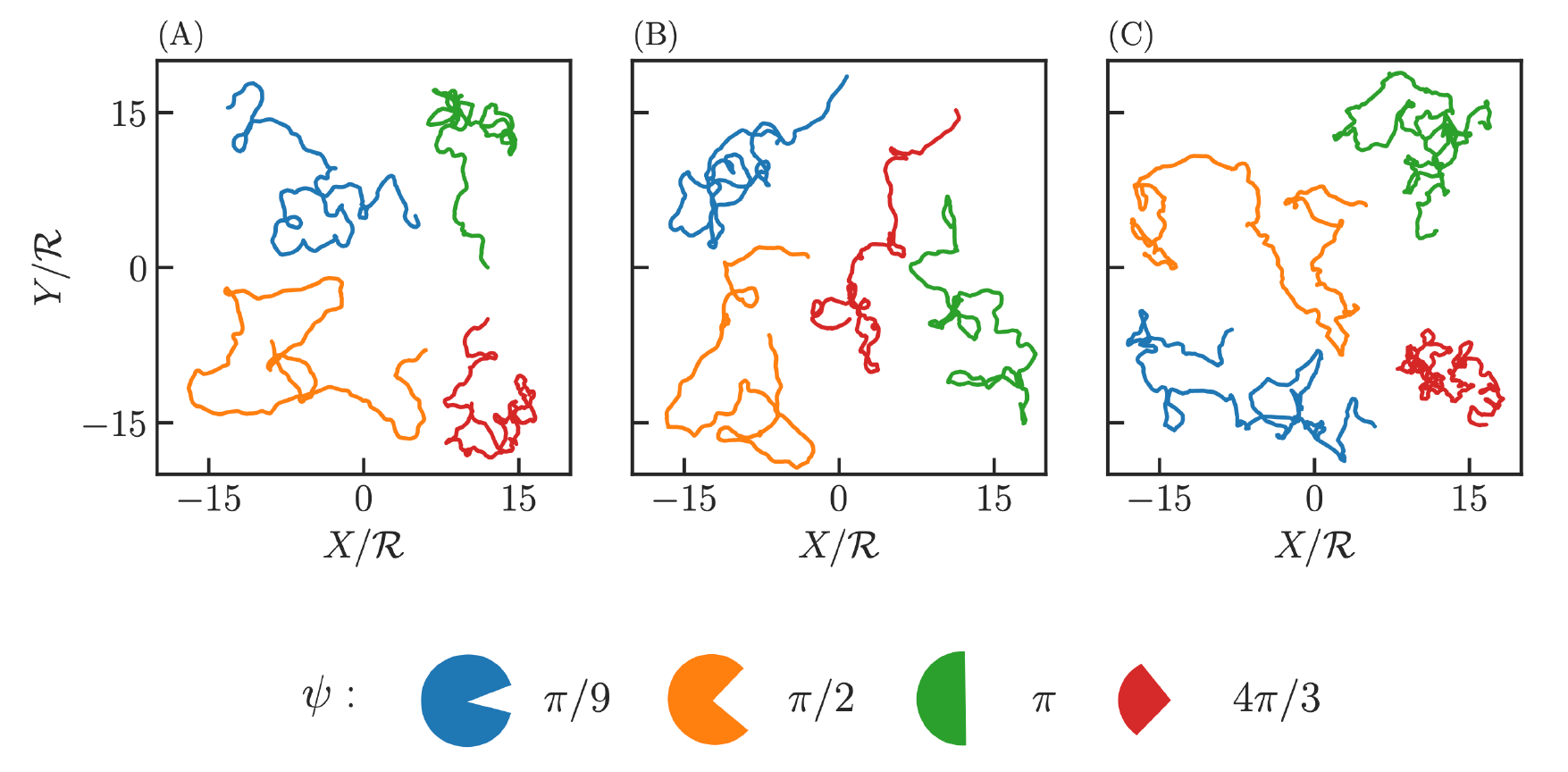}
\caption{Typical trajectories (all of duration $25 \, \omega^{-1}$) of the center-of-mass $\mathbf{R}$  of the inclusion at $\mathrm{Pe} = 66$ and $\phi=0.3$ for various $\psi$ in a bath of (a) RTPs, (b) ABPs and (c) AOUPs. The colors of the trajectories correspond to various values of the opening angle $\psi$, and thus the shape of the inclusion. Notice that the persistent motion of the inclusion scales upto several $\mathcal{R}$. This persistent motion is pronounced when $\psi \sim \pi/2$. See SI Movies.}
\label{fig:trajectories}
\end{figure}

We find that for a large range of parameter values, i.e., the opening angle of the inclusion $\psi$, the area-fraction $\phi$ and the P\'eclet number $\mathrm{Pe}$, the polar inclusions show persistent motion for short times. In FIG.~\ref{fig:trajectories}, we plot the typical trajectories of the center-of-mass of the inclusion $\mathbf{R}$ at three different opening angles $\psi$ for the three models that we study. We notice that for small $\psi$ the trajectories do not display persistent motion. This is also true for large $\psi$. When $\psi \sim \pi/2$, we see long stretches of persistent motion of the inclusion. However, at long times, the movement of the inclusion is diffusive in nature for all angles $\psi$. In other words, a polar inclusion in an active bath shows persistent motion at short times and crosses over to diffusive dynamics at long times -- the inclusion itself behaves as an `emergent active particle'.

\begin{figure}[ht]
\includegraphics[width=\linewidth]{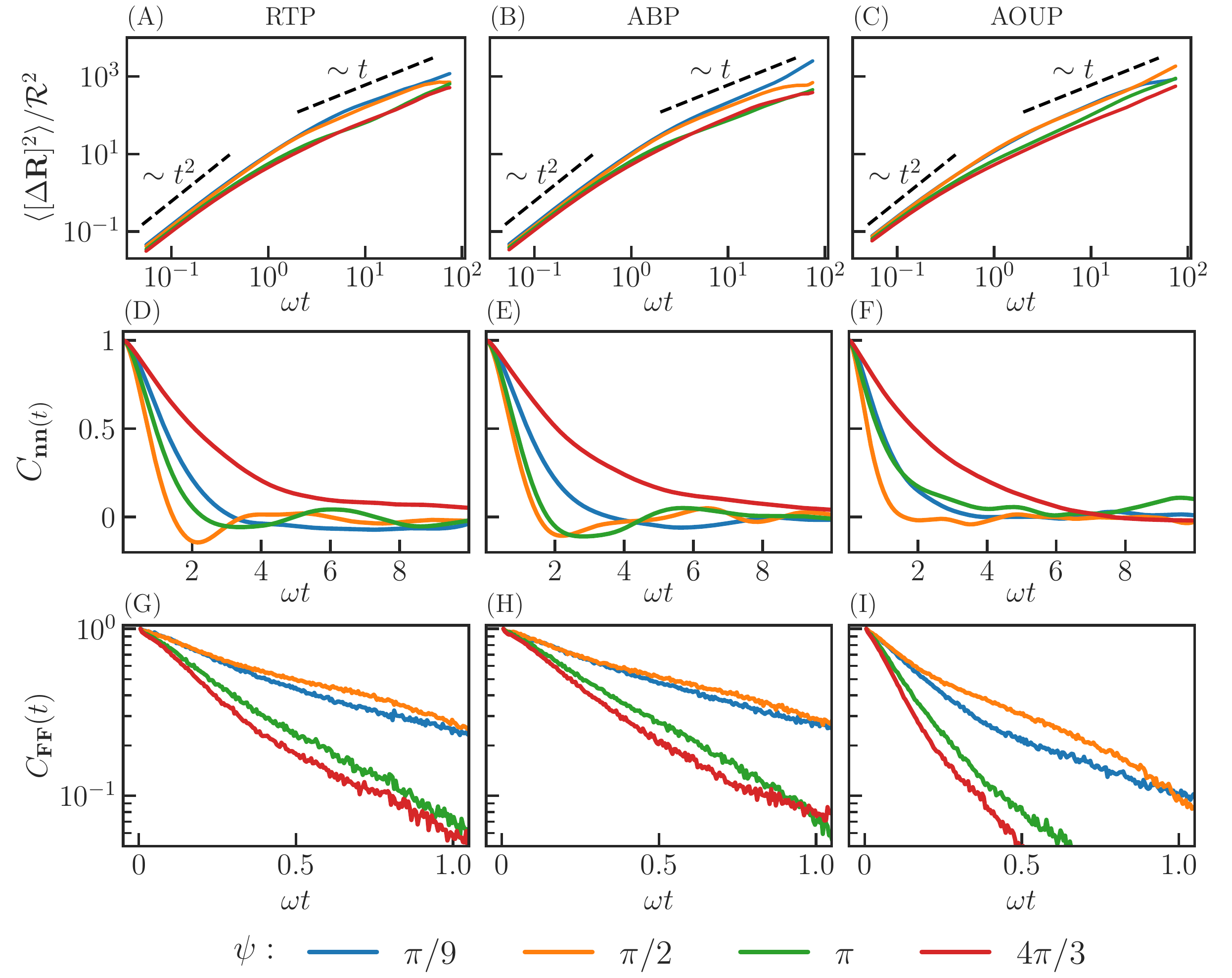}
\caption{Dynamical correlations of the polar inclusion at $\mathrm{Pe}=66$ and $\phi =0.3$. Mean-squared displacement of the inclusion in a bath of (A) RTPs, (B) ABPs, and (C) AOUPs. All models display a transition from an initial ballistic behavior $\sim t^2$ to an asymptotic diffusive behavior $\sim t$ confirming the persistent motion of the inclusion at short times. (D-F) The orientatation correlations as a function of time for three different models. The orientation correlation $C_{\mathrm{\mathbf{n}\mathbf{n}}}$ decays monotonically for the AOUP bath (F) while for (D) RTP  and (E) ABP baths, it has a negative minima around $\omega t \sim 2$ for inclusion shapes that show enhanced persistent motion. The two-point correlation function of the force on the inclusion \eqref{eq:force_correlation} shows approximately an exponential decay in time for all three models (G-I). However, the decay time $\tau_F$ is a non-monotonic function of $\psi$ (see FIG.~\ref{fig:force_correlation_time}).}
\label{fig:pacman_msd_angle_force_correlation}
\end{figure}

To further study this persistent motion of the inclusion, we plot in FIG. \ref{fig:pacman_msd_angle_force_correlation} (A-C) the mean-squared displacement (MSD) $\langle \left[ \Delta \mathbf{R}(t)\right]^2\rangle=\langle \left[ \mathbf{R}(t) - \mathbf{R}(0)\right]^2\rangle $ for the three models and various opening angles $\psi$ at $\phi=0.3$ and $\textrm{Pe}=66$. The MSD shows ballistic behavior $\langle \left[ \Delta \mathbf{R}(t)\right]^2\rangle \sim t^2$ for short times which eventually crosses over to diffusive dynamics $\langle \left[ \Delta \mathbf{R}(t)\right]^2\rangle \sim t$ at times larger than $ \omega^{-1}$. 

An alternate measure of the persistent motion of the inclusion is to calculate the auto-correlation $C_{\mathbf{n}\mathbf{n}}(t) = \langle \hat{\mathbf{n}}(0) \cdot \hat{\mathbf{n}}(t)\rangle$ of the unit-vector $\hat{\mathbf{n}} = -\cos\Phi \, \hat{\mathbf{x}} - \sin\Phi \, \hat{\mathbf{y}}$ (see FIG.~\ref{fig:schematic}). The dependence of $C_{\mathbf{n}\mathbf{n}}(t)$ is shown in FIG.~\ref{fig:pacman_msd_angle_force_correlation} (D-F) for the three active bath models. We observe that $C_{\mathbf{n}\mathbf{n}}$ decays to zero at long times $t \gg  \omega^{-1}$. However, while $C_{\mathbf{n}\mathbf{n}}(t)$ decays monotonically to zero for the AOUP bath, it shows oscillations for the RTP and ABP baths. This is attributed to the different stochastic nature of the active velocity: the magnitude of the active velocity is fixed for RTPs and ABPs whereas for AOUPs it is drawn from a Gaussian distribution with a characteristic value $u$. As such the orientation of the inclusion decorrelates quickly for the AOUP bath. This is also observed in the MSD of the inclusion angle $\langle [\Delta \Phi(t)]^2 \rangle$ where the crossover to the diffusive regime occurs at timescales $t \lesssim 1/\omega$ for AOUP bath whereas it occurs at timescales $t \gtrsim 1/\omega$ for RTP and ABP baths.

The short-time ballistic dynamics of the MSD seen in FIG \ref{fig:pacman_msd_angle_force_correlation} (A-C) and the long-lived two-point correlations of the unit normal vector FIG \ref{fig:pacman_msd_angle_force_correlation} (D-F) clearly indicate persistent motion of the inclusion. However, this behavior is in the configurational degrees of freedom, i.e., $\mathbf{R}$ and $\hat{\mathbf{n}}$, of the inclusion. How does this persistence emerge from the coordinated impacts on the inclusion due to active particles constituting the bath? To answer this question, we look at the force $\mathbf{F}$ on the center-of-mass of the inclusion defined in \eqref{eq:force_on_inclusion}. In the steady-state, the average force vanishes $\langle \mathbf{F}(t) \rangle =\mathbf{0}$. To quantify the temporal dynamics of $\mathbf{F}(t)$, we measured the normalized two-point force correlation function:
\begin{align}
C_{\mathbf{F}\mathbf{F}}(t) = \frac{\langle \mathbf{F}(t) \cdot \mathbf{F}(0)\rangle}{\langle \mathbf{F}(0) \cdot \mathbf{F}(0)\rangle}.
\label{eq:force_correlation}
\end{align}
Figure \ref{fig:pacman_msd_angle_force_correlation} (G-I) shows that $C_{\mathbf{F}\mathbf{F}}(t)$ decays exponentially for all opening angles $\psi$ and the decay time $\tau_{F} \sim \omega^{-1}$. In other words, the dynamical force acting on the inclusion has the characteristics of an exponentially correlated noise \cite{Milos2020}.  The torque $\mathsf{T}$, defined in \eqref{eq:torque_on_inclusion}, on the center-of-mass of the inclusion vanishes on the average. And its two-point correlation function decays exponentially as well. However, we found that the decay time of this torque-torque correlation function is largely independent of $\psi$. This suggests that the dynamics of the orientational degree of the inclusion $\Phi$ does not depend on its shape.

\begin{figure}[t]
\includegraphics[width=\linewidth]{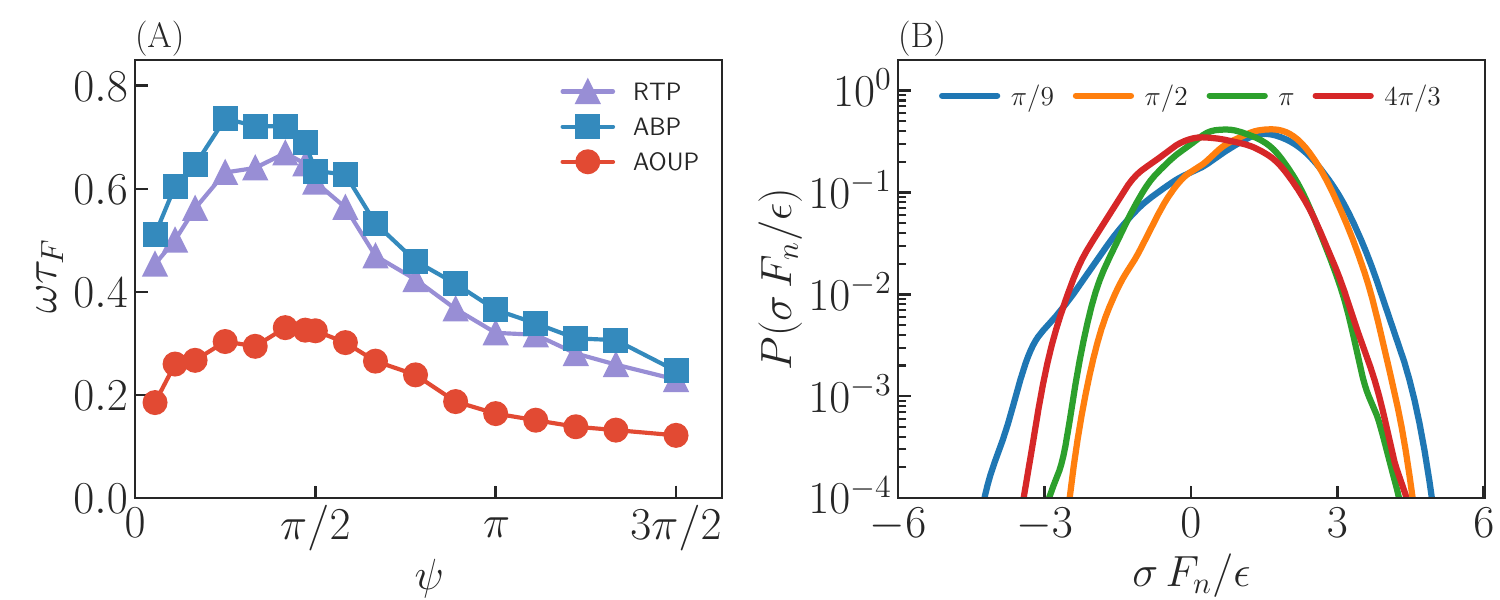}
\caption{(A) The decay time $\tau_F$ of the two-point correlation function $C_{\mathbf{F}\mathbf{F}}(t)$ of the force shows a peak as the shape of the inclusion is changed by varying the opening angle $\psi$. Although the location of the peak coincides for all the three models, the maximal value of $\tau_F$ for the AOUP bath is lower than that for the RTP and ABP baths. (B) The probability distribution of the normal component of the force $F_n = \mathbf{F} \cdot \hat{\mathbf{n}}$ on the inclusion at various values of $\psi$ for the ABP bath. Notice that the location of the peak of this distribution changes with $\psi$ while its width is largely unchanged. We observe a similar result for the RTP and AOUP baths.}
\label{fig:force_correlation_time}
\end{figure}

We note from FIG. \ref{fig:pacman_msd_angle_force_correlation}(G-I) that the decay time $\tau_F$ of $C_{\mathbf{F}\mathbf{F}}(t)$ has a non-monotonic dependence on $\psi$. In fact, $\tau_F$ has a maximum around $\psi \sim \pi/2$ as seen in FIG.~\ref{fig:force_correlation_time}(A). In other words, the persistent active forces on the inclusion from the heat bath depend in a non-trivial way on the shape of the inclusion.  It is important to realize that the inclusion is a polar object for all angles $\psi$. However, for $\psi < \pi$ the region enclosed by the inclusion is a non-convex region, while for $\psi \geq \pi$ the enclosed region is convex. Specifically, the non-convex shapes represent a wedge-like trap that can lead to enhanced accumulation of the active particles in this region thus leading to stronger persistent motion of the inclusion.

\subsection{Shape dependent motility of the polar inclusion}

Though $\langle \mathbf{F} \rangle = \mathbf{0}$ in a fixed frame, the component of $\mathbf{F}$ along the normal to the inclusion defined as $F_n \equiv \mathbf{F} \cdot \hat{\mathbf{n}}$ has a non-zero average. In FIG.~\ref{fig:force_correlation_time}(B), we plot the distribution of $F_n$ for the ABP model at various opening angles $\psi$. We note that while $\langle F_n \rangle$ depends on $\psi$, the variance of the distribution is largely independent of the opening angle of the inclusion. These observations hold for the RTP and AOUP models as well.

\begin{figure}[t]
\includegraphics[width = 0.9\linewidth]{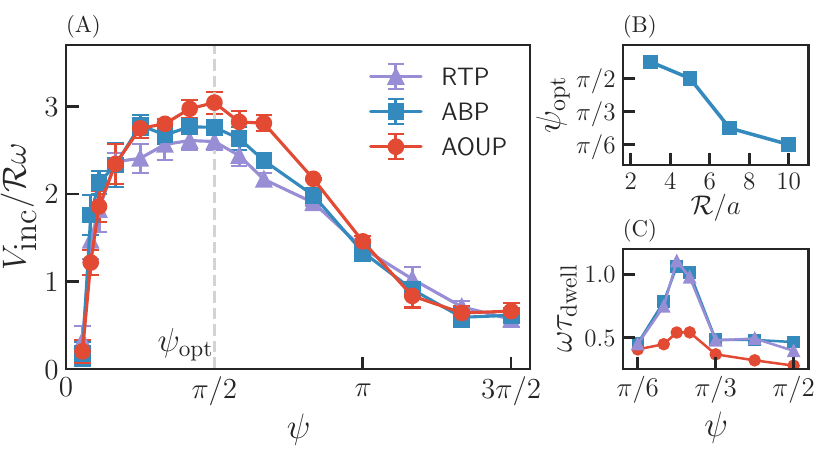}
\caption{(A) The variation of the mean persistent motility of the inclusion defined in \eqref{eq:inclusion_velocity} as a function of its shape. $V_{\mathrm{inc}}$ peaks around $\psi \sim \pi/2$ for all models. Interestingly, the numerical values of $V_{\mathrm{inc}}$ are also similar for all $\psi$ across the three models. Remarkably, we observe that $V_{\mathrm{inc}}$ \emph{does not} change sign as the shape of the inclusion changes from a non-convex region (for $\psi < \pi$) to a convex region (for $\psi > \pi$). In other words, the emergent mean velocity of the anisotropic inclusion is always along $\hat{\mathbf{n}}$. (B) The optimum angle $\psi_{\mathrm{opt}}$ at which $V_{\mathrm{inc}}$ peaks decreases with the radius $\mathcal{R}$ of the inclusion. (C) The time spent by the active particles in the wedge-like trapping region follows an exponential distribution with a characteristic dwell time $\tau_{\mathrm{dwell}}$. This dwell time depends on the shape of the inclusion and shows a peak  around $\psi \sim \pi/4$. The value of $\tau_{\mathrm{dwell}}$ is higher for RTP and ABP baths compared to the AOUP bath. In this plot, $\mathrm{Pe}=66$ and $\phi=0.3$.}
\label{fig:inclusion_velocity}
\end{figure}

Clearly, $F_n$ controls the emergent motility of the inclusion. We define the motility of the inclusion as 
\begin{align}
V_{\textrm{inc}} = \left\langle \frac{d\mathbf{R}}{dt}\cdot \hat{\mathbf{n}} \right\rangle = \mathcal{M}_t \; \left\langle \mathbf{F} \cdot \hat{\mathbf{n}} \right\rangle.
\label{eq:inclusion_velocity}
\end{align}
Note that, in principle $V_{\textrm{inc}}$ can change sign when the shape of the inclusion changes, unlike the active speed $u>0$. In FIG.~\ref{fig:inclusion_velocity}(A), we show the variation of $V_{\textrm{inc}}$ as a function of the opening angle $\psi$ for an area fraction $\phi=0.3$ and $\mathrm{Pe}=66$. It is immediately obvious from the figure that $V_{\textrm{inc}}$ has a maximal value around $\psi_{\textrm{opt}} \sim \pi/2$. Several points are to be noted from FIG.~\ref{fig:inclusion_velocity}(A). First, the variation of $V_{\textrm{inc}}$ as a function of $\psi$ has a similar behavior across all three models. In fact, even the numerical values are also very similar. This trend is consistent with the variation of $\tau_F$ seen in FIG.~\ref{fig:force_correlation_time}(A). Second, $V_{\textrm{inc}}>0$ for all angles $\psi$. This is rather surprising since it is possible that the emergent velocity of the inclusion could change sign with its shape. In particular, for $\psi \leq \pi$, the inclusion has a non-convex shape and changes over to a convex shape for $\psi>\pi$ (see FIG.~\ref{fig:trajectories}). As such, one might have expected a sign reversal of $V_{\textrm{inc}}$ around $\psi\sim\pi$. Our numerical simulations, however, show that this is not the case. Third, as might be expected, $V_{\textrm{inc}}$ vanishes both at $\psi\sim0$ and $\psi\sim2\pi$. For small $\psi$, the opening wedge like region is not wide enough to trap sufficient active particles, while for $\psi\sim2\pi$ the inclusion approaches the shape of a line. In both cases, the inclusion does not have any net polarity in its shape. Fourth, to check how $V_{\textrm{inc}}$ depends on the overall size of the inclusion, we considered the ABP model and varied the radius $\mathcal{R}$. We found that the optimum speed $V_{\rm opt} = V_{\textrm{inc}}(\psi_{\rm opt})$ is largely independent of $\mathcal{R}$. However, the optimal angle $\psi_{\rm opt}$ decreases with $\mathcal{R}$ as shown in FIG.~\ref{fig:inclusion_velocity}(B). Finally, we note that $V_{\textrm{inc}}$ is comparable in magnitude to the coefficient of the ballistic term in the MSD of the inclusion shown in FIG.~\ref{fig:pacman_msd_angle_force_correlation}(A-C).

As remarked above, the accumulation of the bath particles in the wedge-like region has a non-trivial effect on the motility of the inclusion. This wedge-like region exists only for opening angles $\psi < \pi$. The bath particles dynamically enter and leave this region as the inclusion moves around. What is the mean dwell time of the bath particles within this region? The dynamics of a bath particle deep inside this wedge-like region, i.e., $\rho \ll \mathcal{R}$, is hindered by the subsequent accumulation of other active particles at $\rho \lesssim \mathcal{R}$. As such, the dwell time would be an increasing function of $\mathcal{R}$.  At a given $\mathcal{R}$, the opening angle $\psi$ also influences the dwell time with small values of $\psi$ being more effective. However, the impacts due to the small number of particles at small $\psi$ would not lead to a substantial net force on the inclusion. On the other hand, for larger values of $\psi \sim \pi$, the trap does not present a confining region. As such, we expect that there is an optimum opening angle at which this trapping effect is maximal. To quantify this effect, we measured the amount of time $t_{\rm dwell}$ that a particle spends in the wedge-like region. Specifically, we considered a bath particle to be ``trapped'' whenever $\rho/\mathcal{R} < \cos(\psi/2)$ where $\rho$ is its distance from the geometrical center of the inclusion and  $\omega \, t_{\rm dwell} > 1/20$. We find that the dwell times follow an exponential distribution with a characteristic time scale $\tau_{\rm dwell}$. We plot $\tau_{\rm dwell}$ as a function of $\psi$ in FIG.~\ref{fig:inclusion_velocity}(C) and find that there is an optimum opening angle when this characteristic dwell time is a maximum. Thus the anisotropic shape and the wedge-like trapping region of our inclusion serve to trap the bath particles and lead to its emergent shape-dependent motility characteristics.

\begin{figure}[t]
\centering
\includegraphics[width=\linewidth]{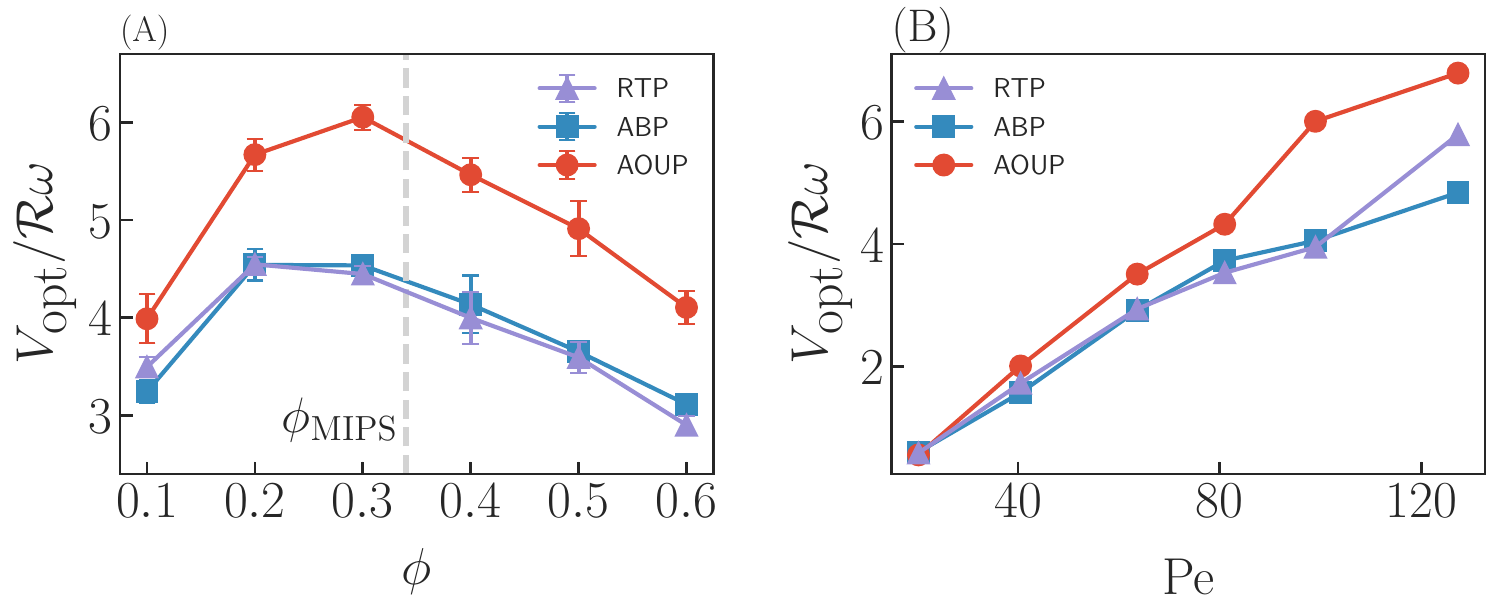}
\caption{(A) The variation of the optimal velocity $V_{\mathrm{opt}}$ (defined as the value of $V_{\mathrm{inc}}$ at the optimal angle $\psi_{\mathrm{opt}}$; see FIG.~\ref{fig:inclusion_velocity}) with the volume fraction $\phi$ at $\mathrm{Pe}=100$. The peak seen in $V_{\mathrm{opt}}$ results from a shielding effect of the active particles on the inclusion (see text). The dashed line denotes the value of $\phi$ at the MIPS phase boundary \cite{rednerStructureDynamicsPhaseSeparating2013}. (B) At $\phi=0.5$, $V_{\mathrm{opt}}$ is a monotonic function of the P\'eclet number $\mathrm{Pe}$.} 
\label{fig:variation_with_phi_Pe_and_trapping}
\end{figure}

To decipher how the characteristics of the active bath controls the motility of the inclusion, we plot the variation of $V_{\textrm{opt}}$ with the volume fraction $\phi$ in FIG.~\ref{fig:variation_with_phi_Pe_and_trapping}(A). We observe that $V_{\textrm{opt}}$ is a non-monotonic function of $\phi$ and has a peak around $\phi=0.3$. How do we understand this optimum volume fraction? At small $\phi$, most of the active particles  accumulate in the wedge-like region of the inclusion. As such, increasing $\phi$ leads to an initial increase in $V_{\textrm{opt}}$. This trend continues until the wedge-like region is completely occupied by the bath-particles. With a further increase in $\phi$, the large number of particles present in the active bath now start clustering around the inclusion in an isotropic manner. In other words, the anisotropic inclusion is increasingly shielded by a cloud of active particles at large $\phi$. This effectively nullifies the shape anisotropy of the inclusion and leads to a decrease in $V_{\textrm{opt}}$. Note that this behavior is displayed by all the three active bath models and, concomittant with the results in FIG.~\ref{fig:inclusion_velocity}, $V_{\textrm{opt}}$ is larger for AOUP baths compared to the RTP and ABP baths.  Thus the volume fraction of the bath has a non-trivial effect on the emergent motility of the inclusion. On the other hand, an increase in the P\'eclet number $\textrm{Pe}$ leads to an expected monotonic increase in $V_{\textrm{opt}}$ as shown in FIG.~\ref{fig:variation_with_phi_Pe_and_trapping}(B).

\section{Discussion}\label{sec4: discussion and conclusion}

We observe the following points from our study. 
First, the emergent motility of the inclusion arises from a combined effect of the dynamics of the inclusion as well as the dynamics of the active particles in the heat baths. In other words, the polar inclusion in our study is \emph{not} a passive tracer \cite{tracerDynamicsMeanField2021, granekAnomalous2022}, nor a stationary trap designed to capture active particles \cite{kaiserHowCaptureActive2012} or traps with prescribed dynamics \cite{kaiserCapturingSelfpropelledParticles2013}. Rather, the degrees of freedom of our inclusion evolve naturally due the impacts of the bath particles. It should also be noted that the net motility of the inclusion is a consequence of the nonequilibrium nature of the heat bath. In an equilibrium heat bath, the inclusion would not display any net motility, even if it had a polar shape.
Second, the active particles in our heat baths are isotropic in shape. This should be contrasted with baths consisting of rod-like active particles. In the case of active baths with polar particles, the orientational degrees of freedom of the bath particles has a significant effect on their trapping dynamics in a static wedge \cite{NkumarPRE2019}.
Third, previous studies have considered wedge-like inclusions (non-closed shape) in active baths and have measured their emergent motility \cite{liaoTransportMovingBarrier2018,kaiserTransportPoweredBacterial2014} . While the emergent dynamics depends on the opening angle of the wedges, the open shape does not allow constructing shapes like that of our inclusion for $\psi > \pi$. As remarked earlier, our polar inclusion always moves along $\hat{\mathbf{n}}$ and does not reverse its direction of motion even when $\psi > \pi$. This is a novel feature of the closed shape of our inclusion.
Fourth, the motility of the inclusion arises from the net force on it due to impacts of the bath particles. This impact force has non-trivial correlations in time \cite{Milos2020, tracerDynamicsMeanField2021}. This is not surprising since the active forces on the bath particles themselves are correlated in time, i.e., they have persistent driving. But what we have demonstrated in our study is that the shape of the inclusion controls, in a non-trivial manner, the temporal correlations of the net force due the bath particles, and thus its emergent motility.
Fifth, concomitant with our observation that increasing $\phi$ beyond a critical value leads to a decrease in $V_{\mathrm{inc}}$ at a fixed $\mathrm{Pe}$, we observe that the emergent motility of the inclusion is optimum at the value of $\phi$ near the MIPS phase boundary \cite{rednerStructureDynamicsPhaseSeparating2013, catesMotilityInducedPhaseSeparation2015}. We notice that below the MIPS transition, the suboptimal accumulation of particles on the inclusion will lead to a decrease in motility. On the other hand, deep in the MIPS regime, enhanced accumulation of the active particles around the inclusion will ``isotropize'' the polar shape of the inclusion and thus reduce its emergent motility. As such, the region around the MIPS phase boundary in the $\mathrm{Pe}-\phi$ plane seems optimal to transduce the forces from the active bath into enhanced motility of the inclusion. However, we do not see a similar peak when $\mathrm{Pe}$ is varied at a fixed area fraction  $\phi=0.5$.
Sixth, all three models of active particle leads to similar emergent dynamics of the inclusion. This is in line with the observation that these models also display universal scaling features in a one-dimensional single file geometry \cite{dolaiUniversalScalingActive2020}.

Our study reveals optimum shapes for inclusions that leads to enhanced motility. These results could potentially have implications for designing the shapes of embedded objects in active baths \cite{reichhardtRatchetEffectsActive2017}. Thus,  an inclusion that controls its shape in a dynamical manner can control its emergent active motility when placed in an otherwise isotropic and unbiased active heat bath. For instance, moving cells or droplets, wherein their shape is controlled by other mechanisms, could display varied motility characteristics in medium wherein the impacts of the bath particles are not governed by a Boltzmann distribution.

In summary, we have studied the dynamics of a polar inclusion in three different active baths, namely ABPs, RTPs and AOUPs. We see that the inclusion behaves as an emergent active particle with short-time ballistic behaviour and long-time diffusive behaviour. This emergent motility arises from the non-trivial correlations of the impact force exerted by the bath particles on the inclusion. These correlations, in turn, are controlled by the shape of the inclusion. In particular, the inclusion (with $\mathcal{R}/a = 5$) has a maximum motility around an opening angle of $\psi \sim \pi/2$. Surprisingly, the inclusion does not reverse the sense of its motion relative to its normal vector even when it changes from a non-convex shape to a convex shape. Remarkably, we find that the inclusion motility peaks close to the volume fraction of the bath particles that corresponds to the onset of MIPS. An emergent shielding effect of the active particles deep in the MIPS regime seems to lead to a reduction in the inclusion motility. It would be interesting to study the emergent interactions between such passive inclusions embedded in active baths to explore the interplay between active fluctuations and the geometry of shape \cite{kaiserMotionTwoMicrowedges2015}. Finally, our predictions can be tested in various experimental systems, for instance, in vibrated granular systems with embedded inclusions.

\section{Acknowledgements}
We acknowledge support of the Department of Atomic Energy, Government of India, under project number RTI4001. PD thanks Arghya Das for many useful discussions. ASR thanks the SN Bhatt summer program of the ICTS-TIFR. KVK's research is supported by the Department of Biotechnology, India, through a Ramalingaswami re-entry fellowship and by the Max Planck Society through a Max-Planck-Partner-Group at ICTS-TIFR. KVK also acknowledges the financial support of the John Templeton Foundation (\#62220). The opinions expressed in this paper are those of the authors and not those of the John Templeton Foundation.

\bibliography{references}

\end{document}